\documentclass[twocolumn]{jpsj3}

\usepackage{times}
\usepackage{graphicx}
\usepackage{color}

\usepackage{array,booktabs}
\usepackage{amsmath}
\usepackage{amssymb}
\usepackage{bm}

\begin{document}

\bmdefine{\ba}{a}
\bmdefine{\bi}{i}
\bmdefine{\br}{r}
\bmdefine{\bk}{k}
\bmdefine{\bq}{q}
\bmdefine{\bqq}{Q}

\title{Microscopic Theory of $\Gamma_3$ Quadrupole Ordering in Pr Compounds \\
on the Basis of a $j$-$j$ Coupling Scheme}

\author{Ryosuke Yamamura and Takashi Hotta}

\inst{Department of Physics, Tokyo Metropolitan University,
Hachioji, Tokyo 192-0397, Japan}

\recdate{\today}

\abst{
Toward the understanding of incommensurate $\Gamma_3$ quadrupole ordering in PrPb$_3$,
we develop a microscopic theory of multipole ordering in $f^2$-electron systems
from an itinerant picture on the basis of a $j$-$j$ coupling scheme.
For this purpose, we introduce the $\Gamma_7$-$\Gamma_8$ Hubbard model
on a simple cubic lattice with the effective interactions 
that induce local $\Gamma_3$ states.
By evaluating multipole susceptibility in a random phase approximation,
we find that the hybridization between $\Gamma_7$ and $\Gamma_8$ orbitals
plays a key role in the emergence of $\Gamma_3$ quadrupole ordering.
We also emphasize that $\Gamma_3$ quadrupole ordering can be understood
from the concept of {\it multipole nesting},
in which the Fermi surface region with large $\Gamma_8$ orbital density
should be nested on the area with a significant $\Gamma_7$ component
when the positions of the Fermi surfaces are shifted by the ordering vector.
This concept cab be intuitively understood 
from the fact that local $\Gamma_3$ doublets are
mainly composed of two singlets between $\Gamma_8$ and $\Gamma_7$ orbitals.
Finally, we discuss the possible relevance of the present theory 
to the experimental results of PrPb$_3$ and point out some future problems 
in this direction of research.
}

\maketitle

\section{Introduction}
\label{sect:introduction}

In recent decades, multipole ordering in $f$-electron systems has attracted
continuous attention in the research field of condensed matter physics.
\cite{Hotta1,Kuramoto-review,Santini,Onimaru-Kusunose}
In general, a multipole is considered to be 
a spin-orbital complex degree of freedom
emerging in a system in which spin and orbital degrees of freedom are tightly
coupled with each other due to a strong spin-orbit interaction.
A description of multipole degrees of freedom has been
provided on the basis of the Stevens' operator-equivalent technique.\cite{Kusunose1}
Among the $f$-electron systems, rare-earth and actinide compounds
with multiple $f$ electrons per ion exhibit diverse multipole phenomena.
In particular, for the case of $n=2$, where $n$ denotes the local
$f$-electron number, intriguing phenomena originating from
non-Kramers degeneracy have been discussed for a long time.
A typical example is considered to be the two-channel Kondo effect,
which is expected to occur in U and Pr ions with the $\Gamma_3$ ground state.
\cite{Cox1,Cox2}
Another example is the modulated antiferro $\Gamma_3$
$O_2^0$ quadrupole ordering observed in PrPb$_3$ with the AuCu$_3$-type
simple cubic structure.\cite{Onimaru1,Onimaru1b}
In this paper, we are interested in the mechanism of this
peculiar quadrupole ordering.

For the investigation of multipole ordering in Pr compounds,
one may think that it is enough to employ an $LS$ coupling scheme
since, in general, $4f$ electrons of Pr$^{3+}$ are considered
to be almost localized.
However, for PrPb$_3$, the situation does not seem to be so simple.
Before the confirmation of the modulated ordering,
the possibility of the antiferro quadrupole ordering has been
discussed.\cite{Tayama,Onimaru2}
Then, the sinusoidal quadrupole ordering has been observed,\cite{Onimaru1,Onimaru1b}
but it seems to be difficult to explain it on the basis of a localized picture.
Furthermore, in PrPb$_3$, the Fermi surfaces have been clearly observed in
a de Haas-van Alphen (dHvA) experiment,\cite{Aoki}
suggesting that it is necessary to consider the quadrupole ordering
from an itinerant picture.
However, it seems to be a difficult task to treat multipole ordering
from a microscopic viewpoint in $f^2$-electron systems.
In fact, the mechanism of the modulated antiferro quadrupole ordering
in PrPb$_3$ has not been clarified yet, although more than ten years
has passed since the discovery of the peculiar quadrupole ordering.

For the explanation of multipole ordering concerning multiple $f$ electrons, 
it seems to be necessary to consider alternative theoretical
research complementary to the $LS$ coupling scheme.
Thus, we believe that it is meaningful to develop a microscopic theory
for multipole ordering from an itinerant picture on the basis of
a $j$-$j$ coupling scheme.\cite{Hotta2,Hotta3,Hotta4}
In fact, recently, several groups have advanced the theoretical investigation
on the multipole ordering in $f$-electron systems
from the microscopic viewpoint by performing first-principles calculations.
\cite{Kotliar,Ikeda1,Ikeda2,Suzuki1,Suzuki2,Suzuki3,Suzuki4}
For the multipole order in CeB$_6$, in which theoretical research based on 
the $LS$ coupling scheme has been carried out for a long time, but recently,
analysis from the itinerant picture has been performed.\cite{Koitzsch}
For the analysis of multipole ordering on the basis of the $j$-$j$ coupling scheme,
we define the multipole operator as the spin-charge density
in the form of a one-body operator.\cite{Hotta2,Hotta3,Hotta4}
Since the $f$-electron state with angular momentum $\ell=3$ contains
seven orbitals, it is desirable to adopt a seven-orbital Hamiltonian
as a realistic $f$-electron model, if possible.
However, such a seven-orbital model is too complicated to be a prototype
to develop the microscopic theory of multipole ordering.
We also encounter difficulties in interpreting the calculation results,
even though we can perform the calculations in the seven-orbital model.
\cite{Hotta4}

Then, we attempt to effectively reduce the number of relevant orbitals.
When we consider the local $f$-electron states on the basis of the $j$-$j$ coupling scheme,
\cite{Hotta5}
we notice that the $\Gamma_3$ ground states are mainly composed of two singlets
among $\Gamma_7$ and $\Gamma_8$ electrons.
\cite{Miyake1,Hotta6,Kubo1,Hattori,Kubo1b,Kubo2}
Note that the double degeneracy originates from the orbital degrees of freedom
in $\Gamma_8$ electrons.
This is consistent with the fact that $\Gamma_3$ is included in the direct products
of $\Gamma_7$ and $\Gamma_8$.
Thus, we discuss
the $\Gamma_3$ quadrupole ordering in $f^2$-electron systems
on the basis of a $\Gamma_7$-$\Gamma_8$ three-orbital Hamiltonian.\cite{Hotta6,Kubo2}
We evaluate the multipole susceptibility of the model in a random phase approximation
(RPA) and attempt to clarify the mechanism of the emergence of incommensurate
quadrupole order from a microscopic viewpoint.

In this paper, we construct the $\Gamma_7$-$\Gamma_8$ three-orbital model
for the $f^2$-electron system from the itinerant picture.
Then, we introduce the effective interactions that stabilize
the local $\Gamma_3$ ground state for the case of $n=2$.
We perform the RPA calculations for the multipole susceptibilities
to discuss the condition for the appearance of quadrupole order.
We emphasize that the hybridization between $\Gamma_7$ and $\Gamma_8$
orbitals is important for the emergence of the $\Gamma_3$ quadrupole ordering. 
We also propose a concept of multipole nesting.
Namely, the Fermi surface region with large $\Gamma_8$ orbital density
is nested on the area with a significant $\Gamma_7$ component
when we shift the positions of Fermi surfaces by the ordering vector.
This is consistent with the fact that local $\Gamma_3$ doublets are mainly
composed of two singlets between $\Gamma_7$ and $\Gamma_8$ orbitals.
Finally, we discuss the possible relevance of the present results to
the incommensurate quadrupole ordering observed in PrPb$_3$
with some comments on future problems.

The paper is organized as follows.
In Sect.~2, the model Hamiltonian with effective interactions
among $f$ electrons is introduced.
After explaining the multipole operators,
we provide the formulation to evaluate multipole susceptibilities in the RPA. 
In Sect.~3, our calculation results on the multipole susceptibilities are shown.
We clarify the key quantity, $\Gamma_7$-$\Gamma_8$ hybridization,
for the emergence of the quadrupole ordering.
We attempt to unveil the microscopic mechanism of incommensurate
quadrupole ordering by focusing on the Fermi surface nesting property
as well as the orbital density distribution on the Fermi surfaces.
In Sect.~4, we provide several comments on the quadrupole ordering
in the present scenario in comparison with the experimental results
observed for PrPb$_3$.
Finally, we summarize this paper.
Throughout this paper, $\hbar=k_{\rm B}=1$.

\section{Model and Formulation}
\label{sect:Formulation}

\subsection{Model Hamiltonian}

To consider the multipole ordering from a microscopic viewpoint
on the basis of the itinerant picture, we set the model Hamiltonian as
\begin{equation}
H = H_{\rm kin} +H_{\rm loc},
\end{equation}
where $H_{\rm kin}$ and $H_{\rm loc}$ denote the kinetic and local terms
for $f$ electrons, respectively.

First let us discuss in detail how to construct the local term $H_{\rm loc}$.
To consider the electronic properties of $f$-electron compounds,
the best way is to treat the seven-orbital model,
including Coulomb interactions,  spin-orbit coupling, and 
crystalline electric field (CEF) potentials.
For instance, the Kondo phenomena have been discussed in detail
on the basis of the seven-orbital Anderson model
hybridized with several conduction bands.
Then, two-channel Kondo effects was found not only in the 
Pr ion but also in Nd and other rare-earth systems in an unbiased manner.
\cite{Hotta7,Hotta8}

Concerning the multipole ordering,
the seven-orbital Hubbard model has also been
analyzed with the use of the RPA
for the evaluation of multipole susceptibility.
Such calculations have actually been performed,\cite{Hotta4}
but it was difficult to clarify the mechanism
of the multipole ordering from a microscopic viewpoint,
mainly due to the complexity originating from the large number of orbitals.
Thus, it is desirable to reduce the number of relevant orbitals
to obtain the effective Hamiltonian for the purpose of grasping 
the essential point concerning the appearance of multipole ordering.

A basic strategy to construct such an effective model is
to exploit a $j$-$j$ coupling scheme.\cite{Hotta5}
As schematically shown in Fig.~1, we first include the effect of
spin-orbit coupling in the one-$f$-electron state characterized
by the orbital $\ell=3$ and spin $s=1/2$, leading to an octet with
a total angular momentum $j=7/2$ and a sextet with $j=5/2$.
Since the energy of the sextet is lower than that of the octet,
we consider the states of the $j=5/2$ sextet to construct
the effective model for rare-earth and actinide compounds for $n \le 6$.
We accommodate $f$ electrons in the level scheme of the one-$f$-electron
states and include the effect of Coulomb interactions among them.

\begin{figure}[t]
\centering
\includegraphics[width=8.0truecm]{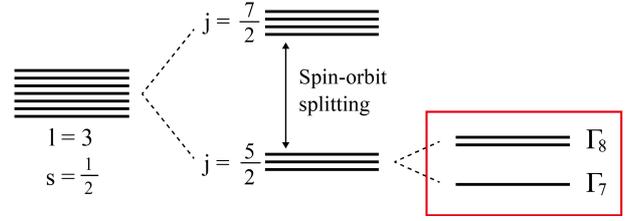}
\caption{(Color online) Level splitting of $f$ electron states
due to spin-orbit coupling, leading to $j=7/2$ octet and $j=5/2$ sextet,
where $j$ is the total angular momentum.
The sextet of $j=5/2$ is further split into a $\Gamma_7$ doublet
and $\Gamma_8$ quartet under the cubic CEF potentials.
Note that for simplicity, we suppress the Kramers degeneracy in this figure.
The effective model is constructed with the use of $\Gamma_7$ and $\Gamma_8$
bases.
}
\end{figure}

Since we consider the cubic system in this paper,
it is convenient to use the cubic irreducible representations.
As shown in Fig.~1, under the cubic CEF potentials,
the $j= 5/2$ sextet is further split into a $\Gamma_7$ doublet
and $\Gamma_8$ quartet.
To distinguish the states in the $\Gamma_8$ quartet and $\Gamma_7$ doublet,
we introduce three pseudo-orbitals $\tau$ (=$a$, $b$, and $c$),
while pseudospin $\sigma$ (=$\uparrow$ and $\downarrow$) is introduced
to distinguish the Kramers degeneracy.
For the description of the model Hamiltonian,
it is useful to define the second-quantized operator $f_{\bi \tau \sigma}$
with pseudospin $\sigma$ and pseudo-orbital $\tau$ at site $\bi$,
expressed as
\begin{equation}
f_{\bi \tau  \sigma}=\sum_{\mu} A_{\tau \sigma,\mu} a_{\bi \mu},
\end{equation}
where $a_{\bi \mu}$ denotes the annihilation operator of an $f$ electron
in the $j=5/2$ sextet with the $z$-component $\mu$ at site $\bi$
and $A$ indicates the coefficient connecting $f$ and $a$ operators.
For $\Gamma_{8a}$ and $\Gamma_{8b}$ electrons,
$f$ operators are explicitly given by
\begin{equation}
\begin{split}
 f_{\bi a  \uparrow}
 &=  \sqrt{\frac{5}{6}}a_{\bi -5/2}+\sqrt{\frac{1}{6}}a_{\bi 3/2},\\
 f_{\bi a \downarrow}
 &= \sqrt{\frac{5}{6}}a_{\bi 5/2}+\sqrt{\frac{1}{6}}a_{\bi -3/2},
\end{split}
\end{equation}
and
\begin{equation}
\begin{split}
 f_{\bi b \uparrow} &=a_{\bi -1/2},\\
 f_{\bi b \downarrow} &=a_{\bi 1/2},
\end{split}
\end{equation}
respectively.

For an $\Gamma_7$ electron ($\tau=c$), we obtain
\begin{equation}
\begin{split}
 f_{\bi c  \uparrow}
 &=  \sqrt{\frac{1}{6}}a_{\bi -5/2}-\sqrt{\frac{5}{6}}a_{\bi 3/2}, \\
 f_{\bi c \downarrow}
 &=  \sqrt{\frac{1}{6}}a_{\bi 5/2}-\sqrt{\frac{5}{6}}a_{\bi -3/2}.
\end{split}
\end{equation}
For the standard time-reversal operator ${\cal K}$=$-{\rm i}\sigma_y K$,
where $K$ denotes an operator taking the complex conjugate,
we can easily show the relation
${\cal K}f_{\bi \tau \sigma}=\sigma f_{\bi \tau -\sigma}$,
which is the same definition for a real spin.\cite{Hotta5}

Now we consider the local term,
which should be composed of the CEF potential and
Coulomb interaction terms.
However, we include only the latter term in this paper
for the reason which we will explain later.
The Coulomb interaction term is given
in the second-quantized form as
\begin{equation}
H_{\rm loc} = \sum_{\bi,1\sim4}
I_{12,34} f^{\dag}_{\bi 1} f^{\dag}_{\bi 2} f_{\bi 3} f_{\bi 4},
\end{equation}
where $I$ indicates the Coulomb interactions and
we use shorthand notation such as $1=\{ \tau_1, \sigma_1 \}$.
Hereafter we use this notation when there is no possibility of confusion.

Concerning the matrix elements of $I$,
there are several methods to evaluate these values.
A straightforward way to obtain $I$ is to estimate the Coulomb integrals
with the use of $f$-electron wavefunctions
in the limit of large spin-orbit coupling $\lambda$.\cite{Hotta5}
This has the merit that we can obtain all the matrix elements analytically,
while we encounter a serious problem that a local $\Gamma_3$ doublet
cannot be stabilized in the limit of infinite $\lambda$.
To reproduce the local $\Gamma_3$ state correctly,
it is necessary to consider the effective interaction, including the effect of
the sixth-order CEF potentials, expressed by the terms of $B_6^0$.\cite{Hutchings}
A simple way to include the effect of $B_6^0$ is to perform
the perturbation expansion in terms of $1/\lambda$
to take into account the effect of the $j=7/2$ octet
in which the $B_6^0$ terms are correctly included.\cite{Hotta9}
Another method is to include the effect of $B_6^0$
through the two-body potential for the $j=5/2$ sextet.\cite{Hotta8}
Finally, it is also possible to more systematically obtain 
the effective interactions to reproduce the low-energy spectrum of
the seven-orbital model by maximizing the overlap integrals
between the states of the seven-orbital model 
and those of the three-orbital model.\cite{Hattori}

In this paper, basically we follow the last method,
but we do not pay special attention to the reproduction of
the low-energy spectrum of the seven-orbital model.
Rather, we simply consider the situation in which
the local $\Gamma_3$ ground state is stabilized
since we are more interested in the mechanism of the $\Gamma_3$
quadrupole order.
An outline of how to determine the effective interaction is as follows.
When we consider local $f^2$ states in the present three-orbital model,
there are 15 eigenstates in total,
originating from a nonet ($J=4$), quintet ($J=2$), and singlet ($J=0)$,
\cite{Hotta5}
where $J$ denotes the total angular momentum of the $f^2$ multiplet
when we suppress the CEF potentials.
Note that the nonet of $J=4$ is the ground-state multiplet
due to Hund's rules.
When we include the cubic CEF potentials, the nonet of $J=4$ is
further split into four groups: 
a $\Gamma_1$ singlet, $\Gamma_3$ doublet, $\Gamma_4$ triplet, and $\Gamma_5$ triplet.
Under the cubic CEF potentials, the quintet of $J=2$ is split into
a $\Gamma_3$ doublet and $\Gamma_5$ triplet.
Namely, the 15 $f^2$ states are classified into
two $\Gamma_1$ singlets, two $\Gamma_3$ doublets,
one $\Gamma_4$ triplet, and two $\Gamma_5$ triplets.
Since the states belonging to the same symmetry are mixed,
in general, the interaction term is expressed as
\begin{equation}
H_{\rm int} = \sum_{\bi,\Gamma,\gamma,p,p'}
V^{\Gamma}_{pp'}
|f^2_{\bi}, \Gamma_{\gamma}^{(p)} \rangle
\langle f^2_{\bi}, \Gamma_\gamma^{(p')}|,
\end{equation}
where $V^{\Gamma}_{pp'}$ indicates the effective interaction,
$|f^2_{\bi}, \Gamma_{\gamma}^{(p)} \rangle$ denotes
the $f^2$ state at site $\bi$,
$\Gamma$ and $\gamma$ characterize the irreducible representation,
and $p$ ($=1$ and $2$) denotes the index used to distinguish
the same irreducible representation.
Note also that $V^{\Gamma}_{pp'}$ does not depend on $\gamma$.

Let us exhibit all the $f^2$ states in the following.
First, for the $\Gamma_1$ singlet states, we obtain
\begin{equation}
\begin{split}
|f^2_{\bi}, \Gamma_1^{(1)} \rangle
&\!=\!\sqrt{\frac{1}{6}} (f^\dagger_{\bi a \uparrow} f^\dagger_{\bi a \downarrow}
+f^\dagger_{\bi b \uparrow} f^\dagger_{\bi b \downarrow}
-2f^\dagger_{\bi c \uparrow} f^\dagger_{\bi c \downarrow})|0\rangle,\\
|f^2_{\bi}, \Gamma_1^{(2)} \rangle
&\!=\!\sqrt{\frac{1}{3}} (f^\dagger_{\bi a \uparrow} f^\dagger_{\bi a \downarrow}
+f^\dagger_{\bi b \uparrow} f^\dagger_{\bi b \downarrow}
+f^\dagger_{\bi c \uparrow} f^\dagger_{\bi c \downarrow})|0\rangle,
\end{split}
\end{equation}
where $|0\rangle$ denotes a vacuum.

\begin{figure}[t]
\centering
\includegraphics[width=8.0truecm]{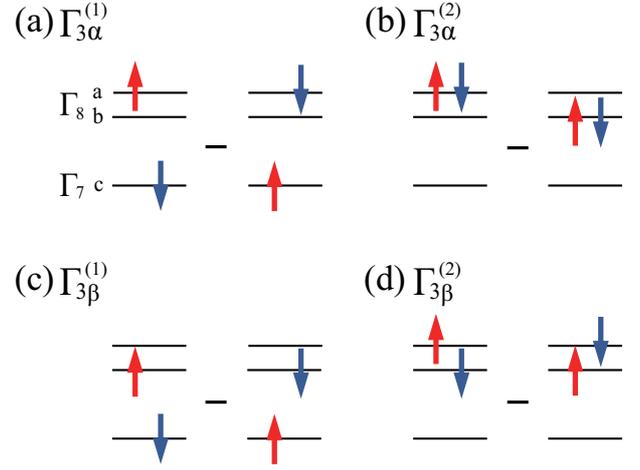}
\caption{(Color online) Local $\Gamma_3$ states composed of two $f$ electrons.
(a) $|f^2_{\bi}, \Gamma_{3\alpha}^{(1)} \rangle$,
(b) $|f^2_{\bi}, \Gamma_{3\alpha}^{(2)} \rangle$,
(c) $|f^2_{\bi}, \Gamma_{3\beta}^{(1)} \rangle$,
and (d) $|f^2_{\bi}, \Gamma_{3\beta}^{(2)} \rangle$.
}
\end{figure}

For the $\Gamma_3$ doublet states,
we introduce $\gamma=\alpha$ and $\beta$ to distinguish the doublet.
For $\Gamma_{3\alpha}$ and  $\Gamma_{3\beta}$, we obtain
\begin{equation}
\label{eq:gamma3a}
\begin{split}
|f^2_{\bi}, \Gamma_{3\alpha}^{(1)} \rangle
&\!=\!\sqrt{\frac{1}{2}}(f^\dagger_{\bi c \downarrow} f^\dagger_{\bi a \uparrow}
- f^\dagger_{\bi c \uparrow}f^\dagger_{\bi a \downarrow})|0\rangle,\\
|f^2_{\bi}, \Gamma_{3\alpha}^{(2)} \rangle
&\!=\!\sqrt{\frac{1}{2}}(f^\dagger_{\bi a \uparrow} f^\dagger_{\bi a \downarrow}
-f^\dagger_{\bi b \uparrow} f^\dagger_{\bi b \downarrow})|0\rangle,
\end{split}
\end{equation}
and
\begin{equation}
\label{eq:gamma3b}
\begin{split}
|f^2_{\bi}, \Gamma_{3\beta}^{(1)} \rangle
&\!=\!\sqrt{\frac{1}{2}} (f^\dagger_{\bi b \uparrow} f^\dagger_{\bi c \downarrow}
-f^\dagger_{\bi b \downarrow} f^\dagger_{\bi c \uparrow})|0\rangle,\\
|f^2_{\bi}, \Gamma_{3\beta}^{(2)} \rangle
&\!=\!\sqrt{\frac{1}{2}} (f^\dagger_{\bi a \uparrow} f^\dagger_{\bi b \downarrow}
-f^\dagger_{\bi a \downarrow} f^\dagger_{\bi b \uparrow})|0\rangle,
\end{split}
\end{equation}
respectively.
As schematically shown in Fig.~2,
$\Gamma_{3\alpha}^{(1)}$ and $\Gamma_{3\beta}^{(1)}$
are given by the singlets between $\Gamma_7$ and $\Gamma_8$ orbitals,
while $\Gamma_{3\alpha}^{(2)}$ and $\Gamma_{3\beta}^{(2)}$
denote the singlets in $\Gamma_8$ orbitals.
Namely, the indexes $\alpha$ and $\beta$ used to distinguish the non-Kramers
$\Gamma_3$ doublets just correspond to $a$ and $b$ orbitals, respectively.
Note that the main components of the $\Gamma_3$ states from $J=4$
($J=2$) are $\Gamma_{3\alpha}^{(1)}$ and $\Gamma_{3\beta}^{(1)}$
($\Gamma_{3\alpha}^{(2)}$ and $\Gamma_{3\beta}^{(2)}$).

For the $\Gamma_4$ triplet state,
we introduce $\gamma=\xi$, $\eta$, and $\zeta$
to distinguish the triple degeneracy, although it is not necessary to introduce $(p)$
since only one $\Gamma_4$ is found.
Then, we obtain
\begin{equation}
\begin{split}
|f^2_{\bi}, \Gamma_{4\xi} \rangle
&\!=\!\left( -\sqrt{\frac{3}{2}} f^\dagger_{\bi b \uparrow} f^\dagger_{\bi c \uparrow}
-\frac{1}{2} f^\dagger_{\bi c \downarrow} f^\dagger_{\bi a \downarrow} \right)|0\rangle,\\
|f^2_{\bi}, \Gamma_{4\eta} \rangle
&\!=\!\sqrt{\frac{1}{2}} (f^\dagger_{\bi c \uparrow} f^\dagger_{\bi a \downarrow}
+f^\dagger_{\bi c \downarrow} f^\dagger_{\bi a \uparrow})|0\rangle,\\
|f^2_{\bi}, \Gamma_{4\zeta} \rangle
&\!=\!\left( -\frac{1}{2} f^\dagger_{\bi c \uparrow} f^\dagger_{\bi a \uparrow}
-\sqrt{\frac{3}{2}} f^\dagger_{\bi b \downarrow} f^\dagger_{\bi c \downarrow} \right)|0\rangle.
\end{split}
\end{equation}

Finally, for the $\Gamma_5$ triplet, we again introduce $\gamma=\xi$, $\eta$,
and $\zeta$ to distinguish the triple degeneracy.
Since we obtain two $\Gamma_5$, it is necessary to prepare $p$ in this case.
Then, we obtain
\begin{equation}
\begin{split}
|f^2_{\bi}, \Gamma_{5\xi}^{(1)} \rangle
&\!=\!\left( \frac{1}{2}  f^\dagger_{\bi b \uparrow} f^\dagger_{\bi c \uparrow}
+\sqrt{\frac{3}{2}}f^\dagger_{\bi c \downarrow} f^\dagger_{\bi a \downarrow} \right)|0\rangle,\\
|f^2_{\bi}, \Gamma_{5\eta}^{(1)} \rangle
&\!=\!\sqrt{\frac{1}{2}} (f^\dagger_{\bi b \uparrow} f^\dagger_{\bi c \downarrow}
+f^\dagger_{\bi b \downarrow} f^\dagger_{\bi c \uparrow})|0\rangle,\\
|f^2_{\bi}, \Gamma_{5\zeta}^{(1)} \rangle
&\!=\!\left( \sqrt{\frac{3}{2}} f^\dagger_{\bi c \uparrow} f^\dagger_{\bi a \uparrow}
-\frac{1}{2} f^\dagger_{\bi b \downarrow} f^\dagger_{\bi c \downarrow} \right)|0\rangle,
\end{split}
\end{equation}
and
\begin{equation}
\begin{split}
|f^2_{\bi}, \Gamma_{5\xi}^{(2)} \rangle
&\!=\! f^\dagger_{\bi a \uparrow} f^\dagger_{\bi b \uparrow} |0\rangle,\\
|f^2_{\bi}, \Gamma_{5\eta}^{(2)} \rangle
&\!=\! \sqrt{\frac{1}{2}} (f^\dagger_{\bi a \uparrow} f^\dagger_{\bi b \downarrow}
+f^\dagger_{\bi a \downarrow} f^\dagger_{\bi b \uparrow})|0\rangle,\\
|f^2_{\bi}, \Gamma_{5\zeta}^{(2)} \rangle
&\!=\! f^\dagger_{\bi a \downarrow} f^\dagger_{\bi b \downarrow} |0\rangle.
\end{split}
\end{equation}

In the present form of the effective interactions,
it is necessary to set 10 parameters for $V^{\Gamma}_{pp'}$.\cite{Hattori}
Namely, there are seven diagonal parameters,
$V^{\Gamma_1}_{11}$, $V^{\Gamma_1}_{22}$,
$V^{\Gamma_3}_{11}$, $V^{\Gamma_3}_{22}$,
$V^{\Gamma_5}_{11}$, $V^{\Gamma_5}_{22}$,
and $V^{\Gamma_4}$,
while we find three off-diagonal parameters,
$V^{\Gamma_1}_{12}$, $V^{\Gamma_3}_{12}$,
and $V^{\Gamma_5}_{12}$.
For simplicity, we define  $v_1=V^{\Gamma_1}_{12}$,
$v_3=V^{\Gamma_3}_{12}$, and $v_5=V^{\Gamma_5}_{12}$.
In this paper, $v_3$ is a control parameter used 
to discuss the quadrupole ordering.

Now we explain the reason why we suppress the one-electron CEF
potential term $H_{\rm CEF}$ in this paper.
Since we use $\Gamma_7$ and $\Gamma_8$ bases in the present model,
the one-electron potential term is given by
$H_{\rm CEF}= b_4 \sum_{\bi}(2\rho_{c\bi}-\rho_{a\bi}-\rho_{b\bi})$,
where $\rho_{\tau\bi}=\sum_{\sigma}f^\dagger_{\bi \tau \sigma}f_{\bi \tau \sigma}$,
$b_4$ is the fourth-order CEF parameter.\cite{Hutchings}
Concerning the experimental finding for the level scheme,
we refer to the result for CePb$_{3}$,
which is the $f^{1}$ compound with the same lattice structure
as that of PrPb$_{3}$.
For CePb$_{3}$, it has been found that $\Gamma_{7}$ is the ground state 
and $\Gamma_{8}$ is the excited state,\cite{Nikl}
indicating that $b_4$ is positive.
Thus, for PrPb$_{3}$, it is recommended to accommodate two $f$ electrons
in this level scheme in the $j$-$j$ coupling scheme.
However, this CEF potential term works toward the destruction of
the $\Gamma_3$ state, which consists of $\Gamma_7$ and $\Gamma_8$ electrons.
We emphasize that our purpose here is to search for the condition for the appearance of
$\Gamma_3$ quadrupole ordering in the $f^2$-electron systems.
Thus, we suppress $H_{\rm CEF}$ from the outset in this paper by taking $b_4=0$,
although we are interested in the effect of the one-electron CEF potential
on the $\Gamma_3$ quadrupole order.

Now we consider the kinetic term.
As emphasized above, in this paper, we discuss the mechanism of
the multipole ordering from the itinerant picture in $f$-electron systems.
For this purpose, we include the hopping of $f$ electrons,
although we do not seriously consider the heavy-mass enhancement
due to the hybridization between localized and conduction electrons.
Then, the kinetic term $H_{\rm kin}$ is expressed as
\begin{equation}
H_{\rm kin} = \sum_{\bk \tau \tau' \sigma }
\varepsilon_{\bk \tau \tau'} f^{\dag}_{\bk \tau \sigma} f_{\bk \tau' \sigma},
\end{equation}
where $f^{\dag}_{\bk \tau \sigma}$ denotes the Fourier transform of
$f^{\dag}_{\bi \tau \sigma}$.
By considering only the hopping between nearest-neighbor sites,
we obtain the one-$f$-electron energy $\varepsilon_{\bk \tau \tau'}$
in a matrix form as
\begin{equation}
{\hat \varepsilon}_{\bk}\!=\!
\begin{pmatrix}
t_{8} \alpha_{\bk} + s_{8} \beta_{\bk} & -\sqrt{3}s_{8} \gamma_{\bk} & t_{78}\beta_{\bk} \\
-\sqrt{3} s_{8} \gamma_{\bk} & t_{8} \alpha_{\bk} - s_{8}\beta_{\bk} & \sqrt{3}t_{78}\gamma_{\bk} \\
t_{78} \beta_{\bk} & \sqrt{3} t_{78}\gamma_{\bk} & t_{7}\alpha_{\bk}
\end{pmatrix},
\end{equation}
where $t_{i}$ and $s_{i}$ denote the hopping amplitudes between adjacent $\Gamma_i$ orbitals,
$t_{78}$ indicates  the hopping amplitude between adjacent $\Gamma_7$ and $\Gamma_8$ orbitals,
$\alpha_{\bk}=\cos{k_x} + \cos{k_y} + \cos{k_z}$,
$\beta_{\bk}=\cos{k_x} + \cos{k_y} - 2 \cos{k_z}$,
and $\gamma_{\bk}= \cos{k_x} - \cos{k_y}$.

Although four hopping amplitudes ($t_7$, $t_8$, $s_8$, and $t_{78}$)
are expressed with the use of four Slater-Koster integrals of
$(ff\sigma)$, $(ff\pi)$, $(ff\delta)$, and $(ff\phi)$,
\cite{Koster,Sharma,Takegahara}
here we use the four hopping amplitudes as parameters for convenience
in this paper.
In the following, we set $t_{8}= -1.0$ and the energy unit is $|t_{8}|$.
Although we do not explicitly mention the chemical potential term in this paper,
the chemical potential $\mu$ is appropriately adjusted under the condition of
$\langle n \rangle=2$,
where $\langle n \rangle$ denotes the average number of $f$ electrons per site.

Finally, we provide a comment on another effective model.
In this paper, we analyze the three-orbital model,
which is composed of $\Gamma_7$ and $\Gamma_8$ orbitals. 
Since the non-Kramers $\Gamma_3$ doublets in the $f^2$ state
are expressed by two local singlets between $\Gamma_7$ and $\Gamma_8$ orbitals,
the $\Gamma_7$-$\Gamma_8$ three-orbital model is suitable for the discussion of 
the multipole ordering in PrPb$_3$ from a microscopic viewpoint.
The $\Gamma_8$ two-orbital Hamiltonian is frequently used as the minimal
model to discuss the multipole ordering in $f$-electron systems,
\cite{Hattori,Shiina1,Shiina2,Kuramoto,Kusunose2,Kubo3,Kubo4,Yamamura}
but the $\Gamma_3$ states composed of a pair of $\Gamma_8$ electrons
are not the main component of the $f^2$ states of the Pr ion.
Thus, we adopt the $\Gamma_7$-$\Gamma_8$ three-orbital model in this paper.

\subsection{Multipole susceptibility in the RPA}

Now we explain the procedure to calculate the multipole susceptibility in the RPA.
First we define the multipole operator in the one-electron-density form
by using the cubic tensor operator $T^{(k)}_{\Gamma \gamma}(\bq)$,
\cite{Hotta2,Hotta3,Hotta4}
where $\bq$ is momentum, $\Gamma$ and $\gamma$ indicate the irreducible
representation for the cubic point group, and $k$ denotes the rank of the multipole.
In the second-quantized form, the cubic tensor operators are given by
\begin{equation}
T^{(k)}_{\Gamma\gamma}(\bq) = \sum_{\tau,\sigma,\tau',\sigma'} 
T^{(K)}_{\tau\sigma,\tau'\sigma'}
f^{\dag}_{\bk \tau \sigma}f_{\bk+\bq \tau'\sigma'},
\end{equation}
where we use the shorthand notation $K=\{k,\Gamma\gamma\}$ and 
the coefficient $T^{(K)}_{\tau\sigma,\tau'\sigma'}$ is given by
\begin{equation}
T^{(K)}_{\tau\sigma,\tau'\sigma'}=
\sum_{\mu,\mu',q} G^{(k)}_{\Gamma\gamma, q}
O^{(k)}_{q,\mu\mu'}A_{\tau\sigma,\mu}A_{\tau'\sigma',\mu'}.
\end{equation}
Here $q$ runs between $-k$ and $k$,
$G^{(k)}_{\Gamma\gamma, q}$ is the transformation matrix
between spherical and cubic harmonics,
and $O^{(k)}_{q}$ denotes the spherical tensor operator
defined in the space of $j=5/2$.
The matrix element of $O^{(k)}_q$ is explicitly calculated by
the Wigner-Eckart theorem as~\cite{Inui}
\begin{equation}
O^{(k)}_{q,\mu\mu'} = \frac{\langle j||O^{(k)}||j \rangle}{\sqrt{2j+1}}
\langle j\mu | j\mu' k q\rangle,
\end{equation}
where $j=5/2$, $\langle j\mu | j\mu' k q \rangle $ indicates
the Clebsch-Gordan coefficient, and $\langle j||O^{(k)}||j \rangle$ is
the reduced matrix element for the spherical tensor operator, given by
\begin{equation}
 \langle j||O^{(k)}||j \rangle=
 \frac{1}{2^k}\sqrt{\frac{(2j + k +1)!}{(2j-1)!}}.
\end{equation}
Note that we define multipole operators from rank 0 to rank 5
in the present model, since the highest rank is given by $2j$. 
Note also that when we express the multipole moment,
we normalize each multipole operator so as to satisfy
the orthonormal condition~\cite{Kubo5}
\begin{equation}
{\rm Tr} \{T^{(K)} T^{(K')} \}=\delta_{KK'}
=\delta_{kk'}\delta_{\Gamma\Gamma'}\delta_{\gamma\gamma'},
\end{equation}
where $\delta$ denotes the Kronecker delta.

\begin{table}
\centering
\begin{tabular}{c|l} \hline
  rank &  irreducible representations \\
\hline
  $0$ & 1g \\
\hline
  $1$ & 4u \\
\hline
  $2$ & 3g, 5g \\
\hline
  $3$ & 2u, 4u, 5u \\
\hline
  $4$ & 1g, 3g, 4g, 5g \\
\hline
  $5$ & 3u, 4u1, 4u2, 5u \\
\hline
\end{tabular}
\caption{Irreducible representations for multipoles up to rank $5$.
Here we use the shorthand notations explained in the main text.
Note that at rank 5, two $\Gamma_{4u}$ triplets appear,
which are distinguished as 4u1 and 4u2.}
\end{table}

In Table~I, we show the list of the irreducible representations
for the possible multipoles up to rank $5$.
Basically, we express the irreducible representations for multipoles
by Bethe notation in this paper,
but we use shorthand notations by combining the number
of irreducible representations and the parity of time-reversal symmetry,
g for gerade and u for ungerade.
For instance, for rank 2, we obtain $\Gamma_{3{\rm g}}$ and $\Gamma_{5{\rm g}}$,
which are simply expressed as ``3g'' and ``5g'', respectively.
Concerning the correspondence to Mulliken notation,
note that $\Gamma_1=A_1$, $\Gamma_2=A_2$, $\Gamma_3=E$, 
$\Gamma_4=T_1$, and $\Gamma_5=T_2$.

In general, it is necessary to consider the linear combination of multipoles.
For instance, multipoles that belong to the same irreducible representation
are allowed to be mixed.
Thus, we introduce the multipole operator $X_{\bq}$
by the linear combination of the cubic tensor operators, given by
\begin{equation}
\label{eq:X}
X_{\bq} = \sum_{k,\Gamma\gamma}P_{K}(\bq)
T^{(k)}_{\Gamma\gamma}(\bq),
\end{equation}
where $P_{K}(\bq)$ indicates the coefficient of the multipole operator
in rank $k$ and irreducible representation $\Gamma\gamma$.

Next it is necessary to consider a way to determine the coefficient $P$.\cite{Hotta4}
For this purpose, we evaluate the multipole susceptibility in the static limit.
In the linear response theory, the multipole susceptibility is defined by
\begin{equation}
\label{eq:chi}
\chi(\bq) = \int^{1/T}_0 d\tau \langle X_{\bq}(\tau)X_{-\bq}(0)\rangle,
\end{equation}
where $T$ is the temperature, 
$X_{\bq}(\tau) = e^{H\tau}X_{\bq}e^{-H\tau}$, 
and $\langle \cdots \rangle$ indicates the thermal average by using $H$.
From Eqs.~(\ref{eq:X}) and (\ref{eq:chi}), we obtain the multipole susceptibility as
\begin{equation}
\chi(\bq) = \sum_{K,K'}P_{K} \chi_{K,K'}(\bq) P_{K'},
\end{equation}
where the susceptibility matrix is given by
\begin{equation}
\label{eq:susmulti}
\chi_{K, K'}(\bq)=
\sum_{1\sim4} T^{(K)}_{1,3} \chi_{12,34}(\bq)T^{(K')}_{2,4}.
\end{equation}
Note that we use the shorthand notations.
We also note that $T^{(K)*}_{i,j}=T^{(K)}_{j,i}$,
where the asterisk denotes the complex conjugate.
Then, $\chi$ and $P$ should be determined by the maximum eigenvalue
$\chi_{\rm max}$ and the corresponding normalized eigenstate of
the susceptibility matrix equation, respectively.

To calculate the multipole susceptibility, in this paper,
we resort to the RPA on the basis of the perturbation expansion
in terms of the Coulomb interactions.
In the RPA, the susceptibility is expressed in a compact matrix form as~\cite{Hattori}
\begin{equation}
\label{eq:rpasus}
{\hat \chi} = {\hat \chi}^{(0)} [{\hat 1}-{\hat I} {\hat \chi}^{(0)}]^{-1},
\end{equation}
where ${\hat 1}$ denotes the unit matrix,
${\hat I}$ is the antisymmetrized interaction in the matrix form,
and the bare susceptibility ${\hat \chi}^{(0)}$ is given by
\begin{equation}
\label{eq:chi01}
\chi^{(0)}_{12,34}(\bq)=
-T\sum_{n,\bk} G^{(0)}_{41}(\bk,i \omega_{n})
G^{(0)}_{32}(\bk + \bq,i\omega_{n}).
\end{equation}
Here $\omega_n=\pi T (2n+1)$ is the fermion Matsubara frequency with integer $n$ and 
$G^{(0)}_{ij}$ is the one-electron Green's function defined from $H_{\rm kin}$.
Note that $G^{(0)}_{ij}=\delta_{\sigma_i,\sigma_j}G^{(0)}_{\tau_i. \tau_j}$.

For actual calculations of ${\hat \chi}^{(0)}$, it is convenient to first diagonalize 
$H_{\rm kin}$ as
\begin{equation}
H_{\rm kin} = \sum_{\bk \nu \sigma}
E_{\bk \nu} {\tilde f}^{\dag}_{\bk \nu \sigma} {\tilde f}_{\bk \nu \sigma},
\end{equation}
where $\nu$ denotes the index used to distinguish the band,
$E_{\bk \nu}$ is the band energy,
and the relation between $f$ and ${\tilde f}$ is expressed as
\begin{equation}
f_{\bk \tau \sigma}=\sum_{\nu} U_{\tau,\nu}(\bk) {\tilde f}_{\bk \nu \sigma}.
\end{equation}
Then, the bare susceptibility is given as
\begin{equation}
\label{eq:chi02}
\begin{split}
 \chi^{(0)}_{12,34}(\bq)
 &\!=\! \delta_{\sigma_1,\sigma_4}\delta_{\sigma_2,\sigma_3}
 \sum_{\bk,\nu,\nu'}
 U^*_{\tau_1,\nu}(\bk)  U_{\tau_4,\nu}(\bk)\\
 &\times  \chi_{\nu, \nu'}(\bk,\bq)
 U^*_{\tau_2,\nu'}(\bk+\bq)  U_{\tau_3,\nu'}(\bk+\bq),
\end{split}
\end{equation}
where $\chi_{\nu,\nu'}(\bk,\bq)$ is given by
\begin{equation}
\chi_{\nu,\nu'}(\bk,\bq)=
\frac{f(E_{\bk+\bq \nu'}) - f(E_{\bk \nu})}{E_{\bk\nu} - E_{\bk+\bq \nu'}},
\end{equation}
and $f$ is the Fermi distribution function.

For the momentum $\bq$ in multipole susceptibility,
we divide the first Brillouin zone into $32 \times 32 \times 32$ meshes.
Namely, the unit of $\bq$ in the present calculation is $\pi/16$.
To efficiently perform the $\bk$ integration in the bare susceptibility Eq.~(\ref{eq:chi02}),
we exploit the Gauss-Legendre quadrature with due care.
First we divide the first Brillouin zone into $16 \times 16  \times 16$ meshes.
Then, in each cube, we adopt $8$-point Gauss-Legendre quadrature
along the $k_x$, $k_y$, and $k_z$ directions.
In this numerical calculation, we can arrive at low temperatures
such as $T/|t_8|=0.01$.

\section{Calculation Results}
\label{Results}

In this section, we show our calculation results for multipole susceptibility.
First we discuss the multipole ordered states
to reveal the condition for the appearance of quadrupole ordering.
Then, we explain that the ordered multipole operators depend on
the local ground states stabilized by the effective interactions among $f$ electrons.
Furthermore, we show that the $\Gamma_3$ quadrupole ordering
is induced by both the  $\Gamma_7$-$\Gamma_8$ hybridization 
and the Fermi surface structure with a nesting property concerning orbital densities.

For the effective interactions, we set
$V^{\Gamma_1}_{11}=0.3$,  $V^{\Gamma_1}_{22}=0.7$,
$V^{\Gamma_3}_{11}=0.0$,  $V^{\Gamma_3}_{22}=0.5$, 
$V^{\Gamma_5}_{11}=0.2$,  $V^{\Gamma_5}_{22}=0.5$,
and $V^{\Gamma_4}=0.1$ for the diagonal parameters.
We again emphasize that here we concentrate only
on the situation with the local $\Gamma_3$ ground state.
For the off-diagonal parameters, we simply set $v_1=v_5=0.1$.
Then, in this paper, we choose the control parameter as $v_3$
to always stabilize the local $\Gamma_3$ ground state.

\subsection{Key role of $\Gamma_7$-$\Gamma_8$ hybridization}

\begin{figure}[t]
\centering
\includegraphics[width=8.2truecm]{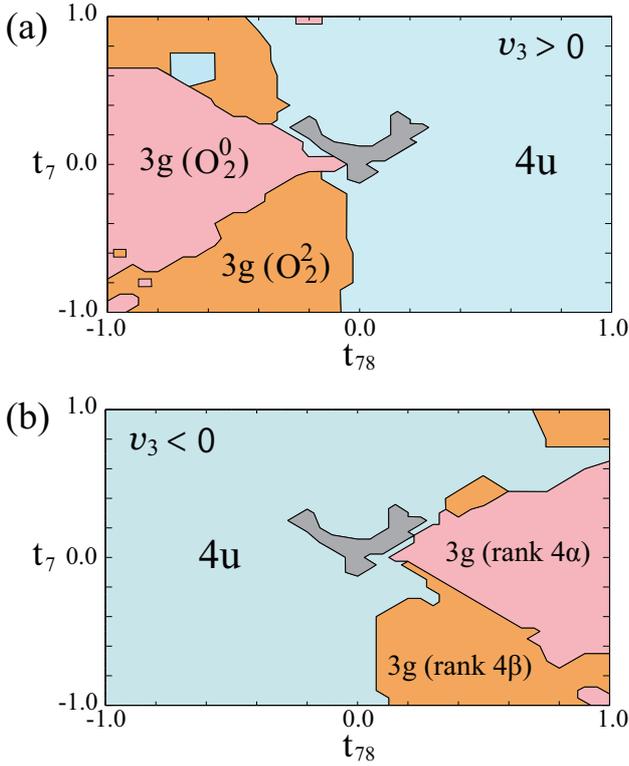}
\caption{(Color online) Phase diagrams of the multipole states for
(a) $v_3>0$ and (b) $v_3<0$
on the $t_{78}$-$t_{7}$ plane for $t_{8}=-1.0$ and $s_{8}=-0.2$.
In the gray region, the RPA susceptibilities already diverge
even at $v_3=0$.
Note that 4u and 3g are distinguished by the multipole susceptibility,
whereas the kind of 3g multipole state is deduced from the main
component in the eigenvector of the multipole susceptibility.
}
\label{phase_dia} 
\end{figure}

First we evaluate multipole susceptibility in the RPA and discuss the multipole phase diagram.
To determine the ordered multipole state and the corresponding $\bq$,
we repeat the calculations of $\chi_{\rm max}$,
the maximum eigenvalue of the RPA susceptibility matrix Eq.~(\ref{eq:rpasus}),
while changing the value of $v_3$.
Note that we consider cases of both $v_3>0$ and $v_3<0$.
In actual calculations, we plot the inverse susceptibility $1/\chi_{\rm max}$
as a function of $v_3$ and find the point $v_3^{\rm c}$
at which $1/\chi_{\rm max}$ crosses the $v_3$ axis
by extrapolation.

Note that the multipole state among the different irreducible representations
is specified by the multipole susceptibility,
but the multipoles belonging to the same irreducible representation
cannot be distinguished in the present calculations of $\chi_{\rm max}$.
In this paper, the kind of multipole in the same irreducible representation
is deduced from the main component in the eigenvector of
the corresponding eigenvalue $\chi_{\rm max}$
at $v_3 = \varepsilon v_3^{\rm c}$, where $\varepsilon$ takes a value
between $0.9$ and $0.98$ depending on the hopping parameters.

In Figs.~3(a) and 3(b), we show the phase diagrams for the ordered multipole states
for $v_3>0$ and $v_3<0$, respectively,
on the $t_{78}$-$t_{7}$ plane for $t_{8}=-1.0$ and $s_{8}=-0.2$.
The 3g multipole states are classified into
$O_2^0$, $O_2^2$, rank $4\alpha$, and rank $4\beta$
from the main component in the eigenvector of
the corresponding eigenvalue $\chi_{\rm max}$
at $v_3 = \varepsilon v_3^{\rm c}$.
Here we focus on this kind of multipole,
while we suppress the information on the ordering vector, 
which will be separately discussed later.
Note that in the present calculations, $v_3=0$ does not mean
the non-interacting case since there are finite other interactions.
In fact, in Figs.~3(a) and 3(b), we find gray regions near $t_{78}=t_7=0$,
in which the RPA susceptibilities already diverge even at $v_3=0$
due to the flat-like $\Gamma_7$ band.
We are not interested in the gray regions
since the magnetic phase is stabilized by an effective interaction other than $v_3$.

In Fig.~3(a), as well as for the gray region,
we find three regions as one magnetic state and
two quadrupole states, $O^0_2$ and $O^2_2$.
For $t_{78}<0$, we mainly obtain the 3g state,
but it is found that 85\% of this 3g state is rank 2 and 15\% is rank 4.
Thus, this state is characterized by the 3g quadrupole $O^0_2$ or $O^2_2$.
Note that the winner of the competition between $O^0_2$ and $O^2_2$
is not determined only by the local conditions, since the local ground states
provide the same contribution to $O^0_2$ and $O^2_2$.
For $t_{78} \ge 0$, magnetic 4u states appear,
which are a mixture of dipoles, octupoles, and dotoriacontapoles.
We remark that the magnetic multipole state always appears
in the present model at $t_{78}=0$.

On the other hand, as shown in Fig.~3(b),
for $v_3<0$, two 3g hexadecapole states are stabilized
in the region of $t_{78}>0$.
In this case, 95\% of the 3g state is rank 4 and 5\% is rank 2,
indicating that the amounts of ranks 2 and 4 are almost reversed
in comparison with the 3g states for $v_3>0$.
Note that ``rank $4\alpha$'' and ``rank $4\beta$'' in Fig.~3(b) indicate
hexadecapoles belonging to the same group as $O^0_2$ ($\Gamma_{3\alpha}$)
and $O_2^2$ ($\Gamma_{3\beta}$), respectively.
For $t_{78} \le 0$, we again find that 4u magnetic multipole states appear.
We suppress the information on the ordering vector $\bqq$ in these diagrams,
where $\bqq$ is defined as the momentum at which the
maximum quadrupole susceptibility appears.
At this stage, we simply comment
that $\bqq$ is different even in the same multipole state,
depending on the hopping amplitudes.

At $t_{78}=0$, we find that the magnetic multipole states always appear
irrespective of the sign of $v_3$.
This is consistent with a previous result.\cite{Kubo2}
Namely, the magnetic ground state in the RPA has also been obtained
in the three-orbital model including only nearest-neighbor hopping
$(ff\sigma)$ and the negative Hund's rule interaction between
$\Gamma_7$ and $\Gamma_8$ electrons.
Note that $t_7=t_{78}=0$ when we consider only $(ff\sigma)$.

Here readers may consider that the above results look strange.
Namely, the electric multipole states are not always stabilized,
in spite of the choice of the interaction parameters with
the local 3g doublet ground state without any dipole moments.
In our calculations, we confirm that the stabilized multipole state is
found to be one of 3g, 4u, and 5u
as long as we change $v_3$ as the control parameter.
Note that the 5u octupole appears only in limited parameter regions
(not shown here) and the competition between 3g and 4u usually occurs
in the present calculations.
The ordered multipole moments and corresponding $\bqq$
also depend on the hopping parameters and other local parameters.
We notice that the multipole phase diagrams for $v_3>0$ and
$v_3<0$ are almost symmetric about the line of $t_{78}=0$.
This tendency is also found when we change hopping parameters
$t_{8}$ and $s_{8}$.
For the stabilization of the 3g quadrupole states,
the condition of $t_{78} \neq 0$ is considered to be important.

We find that in the non-interacting case,
${\rm max}(\chi_{\rm max}^{\rm 4u},  \chi_{\rm max}^{\rm 5u})$
=${\rm max}(\chi_{\rm max}^{\rm 1g},  \chi_{\rm max}^{\rm 3g})$,
where $\chi_{\max}^{\Gamma}$
indicates the maximum eigenvalue of the susceptibility for the multipole $\Gamma$
and ${\rm max}(A,B)$ indicates the larger value of A and B.
The 4u-5u competition as well as the 1g-3g competition in the non-interacting case
depends on the hopping amplitudes.
With increasing $|v_3|$ for both $v_3>0$ and $v_3<0$,
we observe the enhancement of $\chi_{\rm max}^{\rm 3g}$ or $\chi_{\rm max}^{\rm 4u}$.

To understand the competition between magnetic and electric states,
we discuss the low-order terms of the RPA susceptibility,
which provide significant contributions to the difference
between magnetic and electric multipole susceptibilities.
It is possible to roughly sketch the phase diagram
by the evaluation of such perturbation expansion terms,
although the winner of the competition between 3g and 4u is finally determined
by the RPA calculations.

\begin{figure}[t]
\centering
\includegraphics[width=8.0truecm]{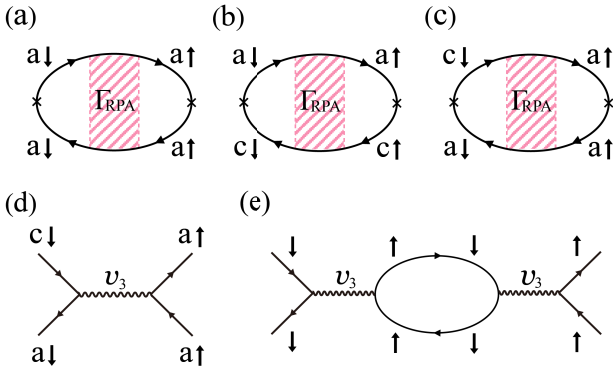}
\caption{(Color online) Feynman diagrams for
(a) $\chi_{a\downarrow a\uparrow, a\downarrow a\uparrow}$,
(b) $\chi_{a\downarrow c\uparrow, c\downarrow a\uparrow}$,
and (c) $\chi_{c\downarrow a\uparrow, a\downarrow a\uparrow}$.
The hatched rectangle $\Gamma_{\rm RPA}$ denotes
the four-point vertex in the RPA.
(d) First- and (e) second-order terms with respect to $v_3$ of $\Gamma_{\rm RPA}$.
}
\label{diagram}
\end{figure}

We analyze the dependence on $\sigma$ and $\tau$ of the RPA susceptibility
by evaluating $\chi_{12,34}(\bq)$ in Eq.~(\ref{eq:susmulti})
for the RPA susceptibility Eq.~(\ref{eq:rpasus}).
After some numerical calculations,
we find that the terms shown in Figs.~4(a)-4(c) induce a significant
difference between magnetic and electric ($\Gamma_{3\alpha}$) susceptibilities.
To estimate the contribution of each Feynman diagram,
we consider the first- and second-order terms of $\Gamma_{\rm RPA}$
in terms of the effective interaction $v_3$.
When we consider the non-interacting case,
the contributions of these Feynman diagrams vanish due to
the conservation of pseudospin
as long as we take into account only the hopping between nearest-neighbor sites.
Namely, the enhancement of such Feynman diagrams strongly depends
on the structure of the four-point vertex $\Gamma_{\rm RPA}$ in the RPA.

In Fig.~4(d), we show the first-order term $\Gamma_{\rm RPA}$
with respect to $v_3$.
Here we remark that the local $\Gamma_3$ states are composed of
two local singlets, as shown in Fig.~2.
Note that $v_3$ is the off-diagonal term between Figs.~2(a) and 2(b)
since we consider the $\Gamma_{3\alpha}$ ($O_2^0$) state.
In Fig.~4(d), the $\Gamma_7$ state appears just once in the scattering process,
leading to the contribution of $G_{ca}G_{aa}^3v_3$ for Fig.~4(a),
$G_{ca}^2G_{ac}G_{aa}v_3$ for Fig.~4(b),
and $G_{cc}G_{aa}^3v_3$ for Fig.~4(c),
where $G_{\tau\tau'}$ denotes the Green's function between
$\tau$ and $\tau'$ orbitals.
Next we consider the second-order terms with respect to $v_3$,
as shown in Fig.~4(e).
Since the bubble includes the Green's functions between different pseudospins,
the second-order terms simply vanish.

When we repeat the above discussion for higher-order terms with respect to $v_3$,
we find that $\Gamma_{\rm RPA}$ never includes the terms of even order of $v_3$.
Thus, concerning the $v_3$ dependence, the contributions of Figs.~4(a)-4(c)
are expressed by an odd function of $v_3$.
Concerning the $t_{78}$ dependence, when we expand
the Green's function $G_{ca}$ in terms of $t_{78}$,
we notice that $G_{ca}$ is an odd function of $t_{78}$.
Thus, the contributions from Figs.~4(a) and 4(b) are given by
an odd function of $t_{78}$,
while that from Fig.~4(c) is considered to be an even function of $t_{78}$.
Then, only for $t_{78} \approx 0$,
the contributions of these Feynman diagrams are almost suppressed,
even though we change the other hopping parameters $t_8$, $s_8$, and $t_7$.
The mechanism can be intuitively understood as follows.

Since the susceptibilities in Figs.~4(a) and 4(b) are considered to be
an odd function of $t_{78}$, they are suppressed
for the case of $t_{78}=0$.
Thus, only the Feynman diagram in Fig.~4(c) remains
since it includes a term independent of $t_{78}$.
Since other effective interactions such as $v_1$ and
$v_5$ are found to enhance the magnetic multipole,
the enhancement of the electric multipoles shown by
the Feynman diagram in Fig.~4(c) is too small to stabilize them,
indicating that the susceptibility of the magnetic multipole is always
larger than that of the electric multipole at $t_{78}=0$.
Namely, for the stabilization of the 3g quadrupole state,
the condition of $t_{78} \neq 0$ is found to be necessary.

From the discussion on the $t_{78}$ and $v_3$ dependences
of the susceptibilities,
it is also possible to qualitatively explain the symmetric behavior of
the multipole phase diagrams shown in Figs.~3(a) and 3(b)
about the line of $t_{78}=0$.
For $t_{78}<0$ and $v_3>0$, the susceptibilities in Figs.~4(a)-4(c) enhance 
the $O^0_2$ quadrupole and suppress the 4u multipole,
leading to the stabilization of the $O^0_2$ state.
On the other hand, for the case of $v_3< 0$,
the $O^0_2$ quadrupole should be suppressed
since these terms are odd functions of $v_3$.
Thus, the 4u multipole is stabilized, 
in sharp contrast to the case of $v_3>0$.
The appearance of the 3g hexadecapole and 4u multipole in the region of
$t_{78}>0$ can also be understood by the dependence of each term
on $t_{78}$ and $v_3$ in addition to the form of the matrix elements of
hexadecapole moments.

\subsection{Fermi surface structure and multipole nesting}

In this subsection, we discuss the ordering vector $\bqq$
of the quadrupole susceptibility when the 3g state is stabilized.
Here we point out an important result.
As shown in Fig.~5, $\bqq$ of the RPA susceptibility is the same as that of
the non-interacting susceptibility if the quadrupole state is stabilized.
Note that we do not mention the comparison between 3g and 4u multipole states.
In the following, we explain this claim in detail.

In Fig.~5, we show Eq.~(\ref{eq:susmulti})
for the RPA susceptibility Eq.~(\ref{eq:rpasus}) and
the bare susceptibility Eq.~(\ref{eq:chi01})
with $K=K'=\{ 2, 3{\rm g}\alpha \}$.
In Figs.~5(a) and 5(b), we show the $O_2^0$ bare and RPA susceptibilities,
respectively, on the $q_y$-$q_z$ plane with $q_x=\pi$
for $t_{8}=-1.0$, $s_{8}=-0.2$, $t_7=0.4$, and $t_{78}=-1.0$.
For these parameters, we obtain the $O_2^0$ quadrupole state
and $\bqq=(\pi, \pi, 5\pi/8)$ in the present RPA calculation.
In actual calculations, we find that $1/\chi_{\rm max}$ becomes
zero at $v_3^{\rm c}=4.3$.
Then, we depict Fig.~5(b) for $v_3=4.0$.
We emphasize that the peak position of the bare susceptibility
in Fig.~5(a) is the same as $\bqq=(\pi, \pi, 5\pi/8)$.

In Figs.~5(c) and 5(d), we show the $O_2^2$ bare and RPA susceptibilities,
respectively, for the same parameters as in Figs.~5(a) and 5(b).
Namely, these are the results in the $O_2^0$ quadrupole state.
The peak position is found to be $\bqq=(\pi, 5\pi/8, \pi)$
for both the bare and RPA susceptibilities.
Since the $O_2^0$ quadrupole ordered state is stabilized,
the magnitude of the RPA susceptibility in Fig.~5(d) is smaller than
that in Fig.~5(b).
It is stressed that the magnitude of the bare susceptibility in Fig.~5(c)
is also smaller than that in Fig.~5(a).
We notice that this tendency is always observed as long as we consider
the quadrupole ordered state in the present research.
Namely, it is possible to deduce the kind of quadrupole
and the ordering vector $\bqq$ in the non-interacting case.

From the analysis of the bare and RPA susceptibilities in the quadrupole state,
we notice that the incommensurability of the quadrupole ordering
is determined by the Fermi surface structure in the non-interacting case,
at least in the RPA.
Concerning the effect beyond the RPA, we provide a comment later.
Then, hereafter we concentrate on the relation between
the Fermi surface structure and the bare susceptibility.

\begin{figure}[t]
\centering
\includegraphics[width=8.0truecm]{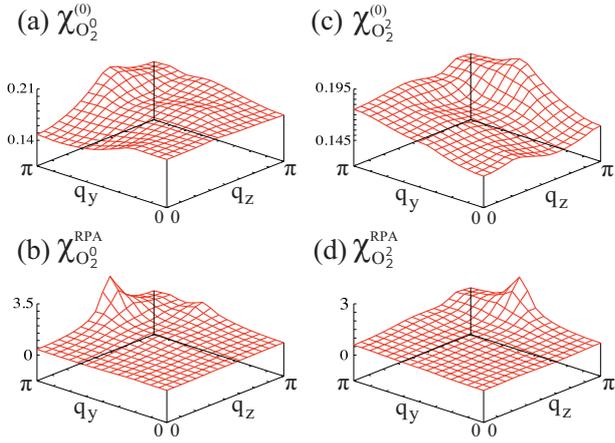}
\caption{(Color online) (a) $O_2^0$ bare susceptibility,
(b) $O_2^0$ RPA susceptibility, (c) $O_2^2$ bare susceptibility,
and (d) $O_2^2$ RPA susceptibility on the $q_y$-$q_z$ plane with $q_x=\pi$
for $t_{8}=-1.0$, $s_{8}=-0.2$, $t_7=0.4$, and $t_{78}=-1.0$.
In the RPA calculation, the $O_2^0$ quadrupole state
is found to be stabilized for these parameters.
To depict (b) and (d), we set $v_3=4.0$.
}
\label{chi_q}
\end{figure}

\begin{figure}[t]
\centering
\includegraphics[width=8.0truecm]{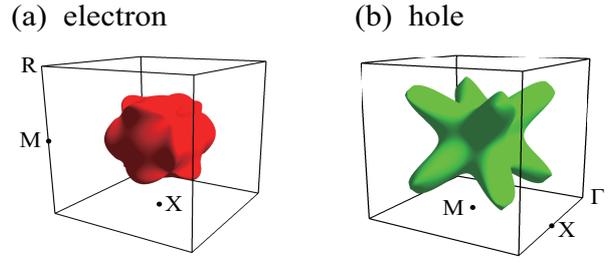}
\caption{(Color online) (a) Electron Fermi surface at the center of the $\Gamma$ point
and (b) hole Fermi surface at the center of the R point
in the present model for $t_{8}=-1.0$, $s_{8}=-0.2$, $t_7=0.4$, and $t_{78}=-1.0$.
}
\label{fermi_3D}
\end{figure}

In Fig.~6, we depict the Fermi surfaces for
$t_{8}=-1.0$, $s_{8}=-0.2$, $t_7=0.4$, and $t_{78}=-1.0$.
The electron Fermi surface in Fig.~6(a) is depicted at the center of the $\Gamma$ point
and is mainly composed of $\Gamma_8$ electrons.
On the other hand, in Fig.~6(b), we show the hole Fermi surface
at the center of the R point, which is composed of $\Gamma_7$ electrons.
Our Fermi surface structure is similar to the result of the band-structure
calculations\cite{Aoki}, except for the small-size Fermi surface, which is not
observed for the present parameters.
When we evaluate $n_7$ and $n_8$, which are the average electron numbers
in the $\Gamma_7$ and $\Gamma_8$ orbitals per ion, respectively,
we obtain $n_7 = 0.79$ and $n_8 = 1.21$ in the present case.
These values seem to be consistent with those expected from
the local $\Gamma_3$ singlets,
although there are deviations from $n_7=n_8=1$
due to the difference in the itinerant properties of
$\Gamma_7$ and $\Gamma_8$ electrons.

When we recall the susceptibility of the one-band model,
$\bqq$ is basically determined from the nesting condition
of the Fermi surface.
We imagine that the nesting is still important for the
determination of $\bqq$ of multipole susceptibility
in the multiband systems,
but it is difficult to conclude the importance of the nesting
only from Fig.~6.
Thus, we analyze the bare susceptibility in more detail.

For this purpose, we perform the multipole decomposition of
the bare susceptibility as in the case of Eq.~(\ref{eq:susmulti}).
After some algebraic calculations, we obtain
\begin{equation}
 \label{eq:chi03}
 \chi^{(0)}_{K,K'}(\bq)=\sum_{\bk}\chi^{(0)}_{K,K'}(\bk,\bq),
\end{equation}
where $\chi^{(0)}_{K,K'}(\bk,\bq)$ is given by
\begin{equation}
\begin{split}
\chi^{(0)}_{K,K'}(\bk, \bq) &=\sum_{\nu\sigma,\nu'\sigma'}
L^{(K)}_{\nu\sigma, \nu'\sigma'}(\bk,\bq)
L^{(K')*}_{\nu\sigma, \nu'\sigma'}(\bk,\bq) \\
&\times \chi_{\nu, \nu'}(\bk,\bq),
\end{split}
\end{equation}
and $L$ is defined as
\begin{equation}
L^{(K)}_{\nu\sigma, \nu'\sigma'}(\bk,\bq)
=\sum_{\tau\tau'}
T^{(K)}_{\tau\sigma,\tau'\sigma'}U^*_{\tau, \nu}(\bk)U_{\tau', \nu'}(\bk+\bq).
\end{equation}

\begin{figure}
\centering
\includegraphics[width=8.0truecm]{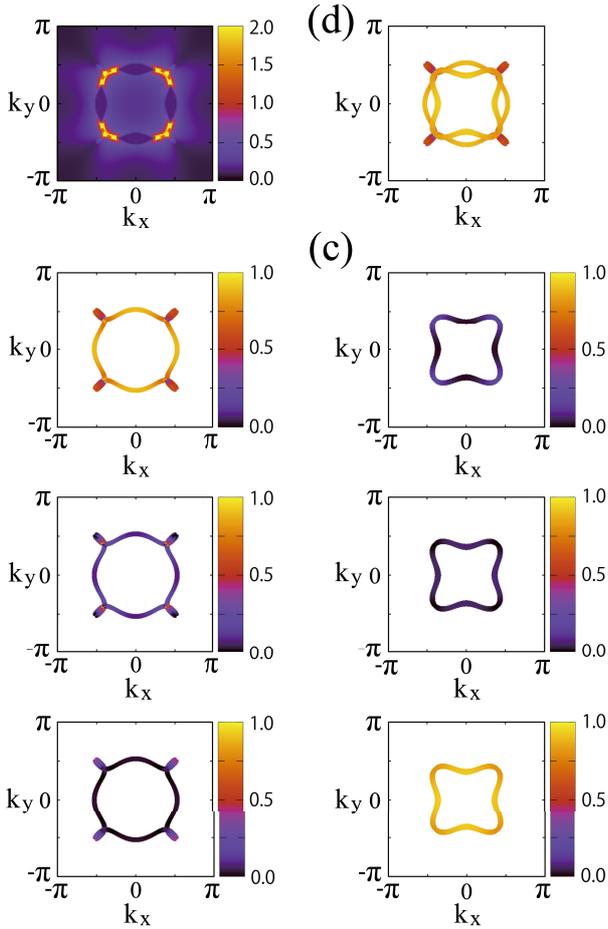}
\caption{(Color online) (a) Color plot of $\chi^{(0)}_{K,K'}(\bk,\bqq)$
for $K=K'=\{ 2, 3{\rm g}\alpha \}$ on the $k_x$-$k_y$ plane for $k_z=3\pi/10$
with $\bqq=(\pi, \pi, 5\pi/8)$.
(b) Orbital densities on the curve defined by $E_{\bk \nu}=\mu$.
(c) Orbital densities on the curve defined by $E_{\bk+\bqq \nu'}=\mu$.
(d) Nesting between $\Gamma_{8a}$ density on the curve of $E_{\bk \nu}=\mu$
and $\Gamma_7$ density on that of $E_{\bk+\bqq \nu'}=\mu$.
}
\end{figure}

In Fig.~7, we show the results for
$t_{8}=-1.0$, $s_{8}=-0.2$, $t_7=0.4$, and $t_{78}=-1.0$.
In this case, we have already found that $\bqq=(\pi, \pi, 5\pi/8)$
in the $O_2^0$ susceptibility.
Then, we focus on the $\bk$ dependence of
$\chi^{(0)}_{K,K'}(\bk,\bqq)$ in Eq.~(\ref{eq:chi03})
for  $K=K'=\{ 2, 3{\rm g}\alpha \}$.
Since it is difficult to depict all the results in the first Brillouin zone,
we exhibit $\chi^{(0)}_{K,K'}(\bk,\bqq)$
on the $k_x$-$k_y$ plane for $k_z=3\pi/10$ in Fig.~7(a).
Note that the value of $k_z$ is chosen for convenience.
The spot-like bright regions denote large contributions to
the susceptibility, but only from this result,
we cannot understand the reason why such regions appear.
Then, we show the orbital densities on the curves defined by
$E_{\bk \nu}=\mu$ and $E_{\bk+\bqq \nu'}=\mu$
in Figs.~7(b) and 7(c), respectively.
We clearly observe that the $\Gamma_{8a}$ and $\Gamma_7$ densities
become significantly large on the curves defined by
$E_{\bk \nu}=\mu$ and $E_{\bk+\bqq \nu'}=\mu$, respectively.

In Fig.~7(d), we consider the nesting between the two curves
$E_{\bk \nu}=\mu$ and $E_{\bk+\bqq \nu'}=\mu$
with significant orbital densities.
Then, we notice the existence of segments on the curves
that satisfy the condition of $E_{\bk \nu}=E_{\bk+\bqq \nu'}$,
leading to the same positions as the spot-like bright regions in Fig.~7(a).
We emphasize the importance of the nesting between
the curve of $E_{\bk \nu}=\mu$ with large $\Gamma_{8a}$ density
and that of $E_{\bk+\bqq \nu'}=\mu$ with large $\Gamma_7$ density.
Namely, for the stabilization of the $O_2^0$ quadrupole order,
it is necessary to obtain the nesting between the Fermi surfaces
with $\Gamma_{8a}$ and $\Gamma_7$ densities.
Thus, we call it {\it multipole nesting} in this paper.

For different hopping parameters,
the $O_2^2$ quadrupole state is found in Fig.~3(a).
We can depict figures similar to Fig.~7, but here
we only explain the difference from Fig.~7
without showing the figures.
In the $O_2^2$ quadrupole state, we observe the
multipole nesting between $\Gamma_{8b}$ and $\Gamma_7$
orbital densities.
In addition, we also find multipole nesting between
$\Gamma_{8a}$ and $\Gamma_{8b}$ orbital densities in this case.
As shown in Eq.~(\ref{eq:gamma3b}), the local $\Gamma_{3\beta}$ doublet
is composed of a pair of singlets, but after some algebraic calculations,
the singlet between $\Gamma_{8b}$ and $\Gamma_7$ orbitals
is found to be the main component of the local $\Gamma_{3\beta}$ doublet.
Another singlet between $\Gamma_{8a}$ and $\Gamma_{8b}$
gives a minor contribution.
This local singlet structure seems to be consistent with the above explanation
for the multipole nesting in the $O_2^2$ quadrupole state.

We provide a comment on the appearance of the 4u magnetic state for $t_{78} \neq 0$.
If the multipole nesting properties for both $O^0_2$ and $O^2_2$ are found to be weak,
the largest susceptibility among the electric multipoles is the 1g hexadecapole
in the non-interacting system.
In this situation, the susceptibilities of the quadrupole become smaller than
that of the magnetic multipole.
Thus, in such a case, the 4u magnetic multipole state is stabilized
for $t_{78} \neq 0$.

From the present calculation results, we propose that
the $O_2^0$ ($O_2^2$) quadrupole ordering is regarded
as the quadrupole density wave state composed of
$\Gamma_{8a}$ ($\Gamma_{8b}$) and $\Gamma_7$ electrons.
To stabilize this state, it is necessary to have nesting between
the segments on the Fermi surface with large $\Gamma_8$ and
$\Gamma_7$ densities.
This is the most important result of this paper.

\section{Discussion and Summary}
\label{sect:discussion and summary}

In this paper, we introduced the $\Gamma_7$-$\Gamma_8$
model Hamiltonian with the effective interactions that
induce the local $\Gamma_3$ ground states for $n=2$.
Then, we estimated the multipole susceptibilities in the RPA
to reveal the condition for the emergence of $\Gamma_3$ quadrupole ordering.
We clarified that the $\Gamma_3$ quadrupole order
can be understood from the concept of multipole nesting,
in which the Fermi surface region with large $\Gamma_8$ orbital density
should be nested on the area with a significant $\Gamma_7$ component
when we shift the positions of the Fermi surfaces with the ordering vector $\bqq$.
This result suggests that the $\Gamma_3$ quadrupole ordering can be 
understood from the combination of
the $\Gamma_7$ and $\Gamma_8$ electrons in the momentum space,
corresponding to the local $\Gamma_3$ doublets composed of two singlets
between $\Gamma_7$ and $\Gamma_8$ orbitals.

In the present work, we proposed that the quadrupole ordering
is regarded as the quadrupole density wave state from the
itinerant picture.
We believe that our scenario works for the understanding of
the $\Gamma_3$ quadrupole order in PrPb$_3$.\cite{Onimaru1}
First we remark that Fermi surfaces have been observed
in PrPb$_3$ in a dHvA experiment.\cite{Aoki}
This fact seems to support the starting point of our present approach
from the itinerant picture for $f$ electrons.
Note, however, that the band-structure calculations were carried out for
LaPb$_3$,\cite{Aoki} not for PrPb$_3$, probably due to the difficulty
in the treatment of non-Kramers $\Gamma_3$ states
from the itinerant picture.
Thus, the contribution of $f$ electrons to the Fermi surfaces seems
to be unclear, but we simply assume that a $4f$-electron admixture should
appear, more or less, in the Fermi surfaces.
We consider that our present model is constructed for such
itinerant $f$ electrons through the hybridization with conduction
electrons.

Next we emphasize that the incommensurate quadrupole order
in PrPb$_3$ is considered to be the sinusoidal wave state.
It is difficult to reproduce such a state from the localized picture.
However, in the itinerant picture, as emphasized in this paper,
it is possible to regard it as the quadrupole density wave state.
Also from this viewpoint, we believe that the present approach
works for PrPb$_3$.

However, there are some problems in the present approach.
One is the incommensurability of the quadrupole order.
The ordering vector of the peculiar incommensurate quadrupole state
has been found to be $\bqq_0=(\pi, \pi \pm \delta, 0)$ and
$(\pi \pm \delta, \pi, 0)$ with $\delta=\pi/4$.
Unfortunately, in the present calculations including only
the nearest-neighbor hopping, we did not reproduce
the quadrupole ordering with $\bqq_0$.
When we include the next-nearest-neighbor hopping and further neighbors,
it may be possible to obtain the quadrupole ordering with $\bqq_0$,
but in the present paper, we did not make such an effort
for the parameter tuning for the reason below.

Another problem relates to the choice of local interactions.
In this paper, to obtain the $\Gamma_3$ quadrupole ordering,
we restricted ourselves only to the situation
in which the local $\Gamma_3$ state
is stabilized by the effective interactions chosen by hand.
We recognize that it is necessary to further investigate the condition
for the effective interactions to obtain the $\Gamma_3$ quadrupole
ordering from the itinerant picture with realistic parameters.

We emphasize that our purpose is to explore a route to
the $\Gamma_3$ quadrupole order from the itinerant picture.
Thus, we did not thoroughly perform the parameter search for
the hopping amplitudes and local interactions within the present model.
Such effort may be a future task, but it is more desirable
to perform the first-principles calculations
to estimate the effective hopping amplitudes and local interactions
with the use of the Wannier basis functions.\cite{Ikeda1}
In the thus obtained three-orbital model, it is highly recommend to
perform the present calculations for the multipole susceptibility in the RPA.
We believe that this is the next step in this direction of research,
when we attempt to further develop the present theory
for the quadrupole ordering in $f^2$-electron systems.

Here we briefly discuss the effect of $H_{\rm CEF}$.
Since it destabilizes the local $\Gamma_3$ states composed of
$\Gamma_7$ and $\Gamma_8$ singlets,
we ignored this term in this paper,
but it may be interesting to consider the quantum critical behavior
induced by CEF potentials in the $\Gamma_3$ quadrupole ordering.
It may be interesting to observe unconventional superconductivity
induced by quadrupole fluctuations near such a critical point.
This is another future issue.

Finally, we provide a brief comment on the determination of $\bqq$
from the interaction viewpoint.
We evaluated the multipole susceptibility in the RPA
in the present paper and arrived at the picture of multipole nesting
for the microscopic understanding of $\Gamma_3$ quadrupole ordering.
For the ordering vector $\bqq$, within the RPA,
we found that $\bqq$ in the RPA susceptibility is the same
as that in the bare susceptibility.
This statement was found to be valid 
when we investigated the present model by using other effective interactions.
It is difficult to prove it mathematically, but we believe that
$\bqq$ of the quadrupole ordering is determined
in the non-interacting case as long as we consider the quasi-particle picture
on the basis of the Fermi liquid theory.
In the perturbation expansion, it is possible to discuss the peak of susceptibility
including the effect of the vertex corrections beyond the RPA.\cite{Ikeda1}
This point is another future problem.

In summary, we discussed the $\Gamma_3$ quadrupole ordering
in $f^2$-electron systems from the microscopic viewpoint.
We emphasized the point that the $\Gamma_3$ quadrupole order
in $f^2$-electron systems can be understood from the multipole nesting,
in which the Fermi surface region with large $\Gamma_8$ orbital density
is nested on the area with a significant $\Gamma_7$ component
when we shift the positions of the Fermi surfaces with the ordering vector.
This is the conceptual finding of the present paper,
although we have not perfectly explained the $\Gamma_3$ quadrupole
order in PrPb$_3$.
The ordering vector will also be explained within the present scheme,
for instance, by evaluating hopping and interaction parameters 
using first-principles calculations,
which is a future task.

\section*{Acknowledgments}

We are grateful to T. Onimaru for the fruitful discussion on the quadrupole ordering in PrPb$_3$.
We also thank K. Hattori and K. Kubo for discussions and comments.
The computation in this work was partly carried out
using the facilities of the Supercomputer Center
of the Institute for Solid State Physics, University of Tokyo.
This work was supported by JSPS KAKENHI Grant Numbers JP16H04017 and JP17J05394.


\end{document}